\documentclass[aoas,nameyear,dvips]{arximspdf}
\usepackage{dcolumn}
\usepackage{graphicx}

\doi{10.1214/09-AOAS304}
\volume{4}
\issue{2}
\pubyear{2010}
\firstpage{743}
\lastpage{763}

\makeatletter
\newtheorem{lemma}{Lemma}
\newcolumntype{d}[1]{D{.}{.}{#1}}
\makeatother

\begin{document}
\begin{frontmatter}

\title{High-throughput data analysis in behavior genetics\protect\thanksref{TG}}
\runtitle{High-throughput data analysis}
\thankstext[*]{TG}{Supported in part by Israel Academy
of Science Grant 915/05.}
\begin{aug}
\author[A]{\fnms{Anat} \snm{Sakov}\corref{}\ead[label=e1]{sakov@post.tau.ac.il}},
\author[B]{\fnms{Ilan} \snm{Golani}},
\author[B]{\fnms{Dina} \snm{Lipkind}} \and
\author[A]{\fnms{Yoav} \snm{Benjamini}}
\runauthor{Sakov, Golani, Lipkind and Benjamini}
\affiliation{Tel Aviv University}

\address[A]{A. Sakov\\
Y. Benjamini\\
Department of Statistics\\
\quad and Operations Research\\
Tel Aviv University\\
Tel Aviv\\
Israel\\
 \printead{e1}} 
\address[B]{I. Golani\\
D. Lipkind\\
Department of Zoology\\
Tel Aviv University\\
Tel Aviv\\
Israel}
\end{aug}

\received{\smonth{11} \syear{2008}}
\revised{\smonth{9} \syear{2009}}

\begin{abstract}
In recent years, a growing need has arisen in different fields for the
development of computational systems for automated analysis of large
amounts of data (high-throughput). Dealing with nonstandard noise
structure and outliers, that could have been detected and corrected in
manual analysis, must now be built into the system with the aid of
robust methods. We discuss such problems and present insights and
solutions in the context of behavior genetics, where data consists of a
time series of locations of a mouse in a circular arena. In order to
estimate the location, velocity and acceleration of the mouse, and
identify stops, we use a nonstandard mix of robust and resistant
methods: LOWESS and repeated running median. In addition, we argue that
protection against small deviations from experimental protocols can be
handled automatically using statistical methods. In our case, it is of
biological interest to measure a rodent's distance from the arena's
wall, but this measure is corrupted if the arena is not a perfect
circle, as required in the protocol. The problem is addressed by
estimating robustly the actual boundary of the arena and its center
using a nonparametric regression quantile of the behavioral data, with
the aid of a fast algorithm developed for that purpose.
\end{abstract}


\end{frontmatter}

\section{Introduction}\label{sec:1}

The open field study of behavior in animals is a subject of interest in
ethology and behavior genetics, and more recently has turned out to be a
working tool in drug discovery and development [Hall (\citeyear{Hall1936});
Bolivar, Cook and Flaherty (\citeyear{BolivarCookFlaherty2000});
Steele et al. (\citeyear{SteeleEtAl2007});
Brunner, Nestlerc and Leahyc (\citeyear{BrunnerNestlercLeahyc2002})]. In such a study an animal is placed in a circular arena,
with no attraction or constraints, and is free to explore it. The
animals behavior is tracked to produce path data: a time series of
recorded locations $(X_{i},Y_{i})$. Typical path data include tens of
thousands of observations per animal with several experimental groups of
animals. Quantitative summaries of the path (known as endpoints), the
simplest example of which is the total distance traveled, are used by
scientists to identify behavioral differences between groups.

Paths generated by rodents in an open field, while seemingly random, are
structured and consist of typical patterns of behavior: progression
segments separated by lingering segments [Drai, Benjamini and Golani
(\citeyear{DraiBenjaminiGolani2000});
Golani, Benjamini and Eilam (\citeyear{GolaniBenjaminiEilam1993})]. The latter are either
complete arrests or segments in time in which the rodent performs small
local movements (e.g., stretching and scanning) which are captured by a
sensitive tracking system.

Path data are prone to suffer from noise and outliers. During
progression a tracking system might lose track of the animal, inserting
(occasionally very large) outliers into the data. During lingering, and
even more so during arrests, outliers are rare, but the recording noise
is large relative to the actual size of the movement (the smallest value
that the noise can take is 1 pixel which ranges between 0.5--2~cm). The
statistical implications are that the two types of behavior require
different degrees of smoothing and resistance. An additional
complication is that the two interchange many times throughout a
session. As a result, the statistical solution adopted needs not only to
smooth the data, but also to recognize, adaptively, when there are
arrests. To the best of our knowledge, no single existing smoothing
technique has yet been able to fulfill this dual task. We elaborate on
the sources of noise, and propose a mix of LOWESS [Cleveland (\citeyear{Cleveland1977})] and
the repeated running median [RRM; Tukey (\citeyear{Tukey1977})] to cope with these
challenges (Section~\ref{sec:2}). Once the path has been smoothed, the
quantitative summaries are computed from the smoothed path data for each
animal, an approach advocated by Ramsay and Silverman (\citeyear{RamsaySilverman1997}).

One of our experiments was conducted in 3 laboratories simultaneously,
and we noticed that measures relating to distance from the wall
[believed to reflect the level of anxiety of a mouse; Hall (\citeyear{Hall1936});
Archer (\citeyear{Archer1973});
Walsh and Cummins (\citeyear{WalshCummins1976});
Finn, Rutledge-Gorman and Crabbe (\citeyear{FinnRutledgegormanCrabbe2003})] were inconsistent across the laboratories. This is known as the
\textit{replicability problem} and is of deep concern in behavioral
research because such experiments are conducted in many laboratories
[Crabbe, Wahsten and Dudek (\citeyear{CrabbeWahlstenDudek1999})]. A close inspection revealed that
although the three arenas were supposed to be circular, one arena was
slightly distorted at a level hardly noticeable to the eye, affecting
the measures related to distance from the wall. Since the actual
location of the wall was not available from the tracking system, the
distances from the wall were computed using the planned center and
radius. Clearly with such a practice, a distorted circular shape leads
to wrong distance computations and consequently harms replicability.

One solution would be to build a new arena, of an exact circular shape,
and rerun the experiment. However, assuring perfect circularity is
difficult, and furthermore, it would not solve the possible imperfect
circularity problem in other laboratories. We offer a solution utilizing
the fact that mice tend to move along the boundary, and use mouse
location within the arena to estimate the position of its wall by a
nonparametric regression quantile [Koenker (\citeyear{Koenker2005})]. The rationale for
the solution proposed and a technique to estimate the arena's center are
presented in Section~\ref{sec:3}.

As noted before, studies of open field behavior may have several
hundreds of animals per study. Hence, any solution needs to be automatic
(e.g., identification of outliers) and fast (so-called
``high-throughput''). Both LOWESS and RRM meet these criteria. Our
experience has shown that embedding existing nonparametric regression
quantile algorithms into a high-throughput environment is difficult due
to their execution time and convergence problems. As a result, we
developed a fast algorithm for that purpose. The algorithm is presented
in Section~\ref{sec:3}, as well as a comparison of its performance with an
existing algorithm.

The motivation and characteristics of the problems addressed in the
paper came from studies of mice, but the statistical and computational
issues are of broader relevance. There are many other examples of
studies involving automatic path tracking, including those of flies
[Branson et al. (\citeyear{BransonEtAl2009});
Valente, Golani and Mitra (\citeyear{ValenteGolaniMitra2007});
Besson and Martin (\citeyear{BessonMartin2005})], pigs
[Lind et al. (\citeyear{LindEtAl2005})], fish and larger marine
animals [Royer and Lutcavage (\citeyear{RoyerLutcavage2008})] and even human babies
[Vitelson (\citeyear{Vitelson2005})], to name a few. Although some of these studies address, in
particular, the complications involved in the analysis of the tracked
path [Lind et al. (\citeyear{LindEtAl2005}); Royer and Lutcavage (\citeyear{RoyerLutcavage2008})], most users of tracking systems
are typically unaware of the consequences of the inherent noise and
outliers, and the burden of providing sufficient protection is shifted
onto the developers of the systems [e.g., the
Ehto-Vision\tsup{\textregistered} tracking system;
Noldus, Spink and Tegelenbosch (\citeyear{NoldusSpinkTegelenbosch2001});
Spink et al. (\citeyear{SpinkEtAl2001})].

The problem of boundary and center estimation of a circle has also a
broader importance and applications in areas such as image processing
[Shapiro (\citeyear{Shapiro1978});
Kim (\citeyear{Kim1984})], physics
[Karimaki (\citeyear{Karimaki1991})] and the analysis
of data gathered from a circular system such as an eye
[Wang, Sung and Venkateswarlu (\citeyear{WangSungVenkateswarlu2005})], to name a few. The common assumption in these
studies is that the circle is perfect and the estimation of the boundary
is reduced to that of the estimation of the radius and the center, which
is mostly done by least squares or maximum likelihood approaches.
Efforts have been devoted to study properties of the estimator and to
develop simple algorithms to solve these nonlinear problems
[Chan, Elhalwagy and Thomas (\citeyear{ChanElhalwagyThomas2002});
Zelniker and Clarkson (\citeyear{ZelnikerClarkson2006})]. Here, too,
we push the current methodology forward by addressing both the
estimation of the center and the boundary when the shape is only
approximately circular. It is our impression that more involvement of
statisticians is needed as statistical issues are ignored or handled
inappropriately, for both tracking and boundary estimation problems.

\section{Smoothing and identification of arrests}\label{sec:2}

\subsection{Noise in the tracking system}\label{sec:2.1}

Let $(X_{i}^{0},Y_{i}^{0})$ be the actual time series of locations, and
$(X_{i},Y_{i})$ the recorded time series, for $i = 1,2,\ldots.$ We
assume
$X_{i} = X_{i}^{0} + \varepsilon_{i}$ and\vspace*{-1pt}
$Y_{i} = Y_{i}^{0} + \delta_{i}$.
The velocities in the two
directions are $V_{X,i}^{0}$ and $V_{Y,i}^{0}$, and the speed is
$V_{i}^{0} = \sqrt{(V_{X,i}^{0})^{2} + (V_{Y,i}^{0})^{2}}$.

\begin{figure}

\includegraphics{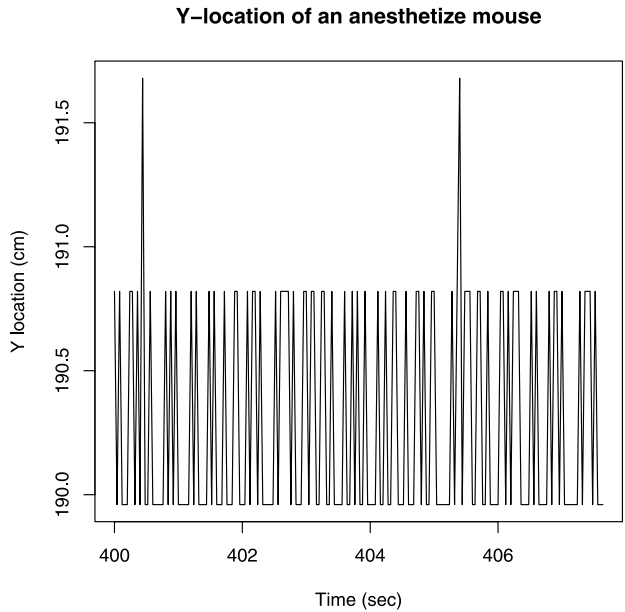}

\caption{A typical 6 seconds of the recorded Y coordinates of
an anesthetized mouse.}\label{fig1}
\end{figure}

There are (at least) three sources for $\varepsilon,\delta $, the first
two are due to the recording noise:
\begin{enumerate}
\item The digital recording of location in systems such as
Etho-Vision\tsup{\textregistered} [Noldus, Spink and Tegelenbosch (\citeyear{NoldusSpinkTegelenbosch2001})] together with the
limited resolution implies that the arena is practically paved with
``tiles'' (in our case they are of size 0.5--2~cm square). In each frame the
system computes the geometrical center of the mouse, and the recorded
location is the center of the tile on which the geometrical center is
found. Since a mouse is larger than one ``tile,'' recordings might
vacillate between neighboring tiles. We call this the precision noise.
Figure~\ref{fig1} illustrates vacillations between two neighboring pixels of the
$Y$-location of an anesthetized mouse, over a few seconds. Naive
computation of the distance traveled by this mouse during a 15~minute
session gives 94 meters.

\item The erratic behavior of the tracking system when it loses track of
the animal inserts outliers that may be large. To assess the extent of
the problem, 30~minute sessions of mice from three strains were
analyzed. We considered an observation to be an outlier if the residual
between recorded and smoothed location was larger than 6 times the
median of the absolute values of the residuals in the window. Slightly
more than 4\% of all observations were outliers.

\item Body wobble consists of movements of the animal which are not part of
its whole-body progression, for example, head scanning or incipient
sideways shifts of weight while running. Although they are real
movements, for the purpose of studying path and velocity they are
unwanted side effects and should be treated as another source of noise.
Their magnitude is different for each animal type, for example, its
magnitude for a turtle is larger than that for a mouse. Hence, heavier
smoothing is needed for a turtle.
\end{enumerate}

Precision noise and body wobble are the main sources of recording noise
during lingering and arrest, while outliers are the main source of
recording noise during progression.

The above examples, as well as the examples presented elsewhere in this
paper, are based on a setup where the arena was circular with radius 125~cm, tracking was performed with the
Ehto-Vision\tsup{\textregistered} tracking
system [Noldus, Spink and Tegelenbosch (\citeyear{NoldusSpinkTegelenbosch2001});
Spink et al. (\citeyear{SpinkEtAl2001})] and recording was at a
rate of 25 or 30 frames per second, for 30~minutes.

\subsection{Smoothing locations and estimating velocities and speed}\label{sec:2.2}

Clearly, a robust smoothing method with smooth derivatives and an
automatic detection of outliers is needed. LOWESS [Cleveland (\citeyear{Cleveland1977})] is
a natural candidate for this purpose. Using a second-degree polynomial,
the locations, velocities and accelerations are estimated for each
direction, as a function of time. We assume that the path, at a small
time window, can be approximated by
\[
X_{i + t} = a_{i} + b_{i}t + c_{i}t^{2} + \varepsilon_{i}, \qquad t = - h, - h + 1, \ldots,0, \ldots,h.
\]
The parameters $a_{i}$, $b_{i}$, $c_{i}$ are estimated using LOWESS to
produce $\hat{a}_{i}$, $\hat{b}_{i}$, $\hat{c}_{i}$. In common applications of
LOWESS interest lies only in the estimation of $a_{i}$ which is the
expectation of $X_{i}$. Here, we are also interested in the velocity and
the acceleration and we make use of all 3 estimated parameters:
\[
\hat{X}_{i} = \hat{a}_{i},\qquad
\hat{V}_{i}^{X} = \hat{b}_{i},\qquad
\hat{A}_{i}^{X} = 2\hat{c}_{i}.
\]
The three quantities, in the $Y$-direction, are found similarly. We
combine the two estimated series of velocities to obtain the time series
of speeds:
$\hat{V}_{i} = \sqrt{(\hat{V}_{i}^{X})^{2} + (\hat{V}_{i}^{Y})^{2}}$.

The data is equally spaced over time, hence, the width of the window is
fixed. We choose a half-window of 10 frames (0.4 seconds), which amounts
to 0.02\% of the data. This is much smaller than the default of Splus or
R, for example. The choice was made by the statisticians and the
biologists involved who compared the smoothed path with the actual
sessions on video, and checked for agreement between them. We also
address this issue in the discussion.

\subsection{Identifying arrests}\label{sec:2.3}

Define an arrest as a period of time, $T$, for which $X_{t} = x$, $Y_{t} =y$ or, equivalently,
$V_{t} = 0$ for $t \in T$. Identifying arrests by
means of zero speed is problematic, as the errors in speeds (compared to
a zero speed) are all positive. Identifying arrests using LOWESS or any
other averaging-based method is problematic since the smoothed locations
are rarely constant due to the averaging nature of these techniques.

The running median, an old yet rarely used method, is appropriate for
the purpose of identifying the \textbf{time} segment of an arrest. In
the repeated running median [RRM; Tukey (\citeyear{Tukey1977})], the running median is
applied iteratively, until convergence, to the sequence obtained in the
previous step. Tukey proposed to perform splitting after convergence.
For computational efficiency we use a variation on that in which we
apply the running median 4 times, iteratively, with half-window sizes of
3, 2, 1 and 1 frames. This is done, separately, for each direction and
an arrest is declared when there is no change in the smoothed locations,
in both directions, for at least 0.2 seconds.

The choice of parameters was tested by comparing arrests found by the
above method with arrests detected by an experienced biologist, watching
videotaped sessions. A 5~minutes session of a mouse of the strain FVB
was taken. The number of arrests found by the moving average, LOWESS and
local polynomial were 40, 25 and 29, respectively, while our method
found 97 arrests. An experienced biologist, blinded by these results,
was asked to count manually the number of arrests she sees. In the
course of 3 repetitions, she got 89, 96 and 102 arrests. The result of
our algorithm is well in the range of arrests counted, while other
methods missed many arrests. Note that even an experienced biologist may
face difficulties in counting arrests (as some are very brief). The
variability would likely be higher if the task was performed by several
biologists. This demonstrates the need for an automated method for
identifying arrests.

\subsection{The combined path smoother}\label{sec:2.4}

The dual task is smoothing location and identifying arrests, when the
two modes of behavior have different characteristics and they
interchange.

LOWESS is not appropriate for identifying arrests due to its averaging
nature. On the other hand, neither the running median nor the RRM is
appropriate for smoothing locations and obtaining velocities and
accelerations, since the resulting path is too rough to represent an
actual movement, and both do not provide (smooth) estimates of
derivatives. Even hanning, which creates a visually more appealing
smooth function, does not help here. See Section~\ref{sec:2.5} for demonstration
of these points.

We are not aware of a single method which addresses both the challenges
of smoothing locations while preserving even short true bursts of
arrests. We find the following combination of LOWESS and RRM to be a
good solution:
\begin{enumerate}
\item Apply LOWESS for each direction to estimate locations and
velocities.
\item Apply the variation of RRM on the \textit{raw} data for
each direction to identify time segments of arrests.
\item When an arrest is found, the velocities in the corresponding
time segments are set to 0.
\item The smoothed locations corresponding to an arrest are linearly
interpolated between the first and last frames of the arrest.
\end{enumerate}

Some samples of the results obtained using the combined procedure can be
viewed in Hen et al. (\citeyear{HenEtAl2004}).

Biologically, arrests and local movements (e.g., head shifts) are
similar, but the latter might look like progression due to the
sensitivity of the tracking system. Once arrests are found, small local
movements should be merged with arrests to create lingering segments.
For that purpose, the maximal speed in all nonarrests segments is
computed and the classification as lingering or progression is described
in Drai, Benjamini and Golani (\citeyear{DraiBenjaminiGolani2000}).

\subsection{Evaluation of the combined path smoother}\label{sec:2.5}

To evaluate the performance of the combined path smoother, we apply the
method to location data of an anesthetized mouse and to simulated paths.
In all cases considered the true location is known. Therefore, the
properties of the smoothed paths can be compared with the properties of
the actual paths.

In the case of the anesthetized mouse, noise comes from the tracking
system itself, and in the simulated paths noise and outliers were built
into the simulation, as described below.

We compare several smoothing approaches on location data of an
anesthetized mouse (that did not move at all) which was tracked for 15
minutes. Using the recorded locations, the distance ``traveled'' is almost
94 m. Using the moving average, local polynomials and LOWESS, prior to
computing distance, produce distances of about 8~m, 13~m and 13 meters,
respectively. Using the combined method, the distance is reduced to
about 3~m, much closer to the true distance which is 0. Moreover, the
difference between the methods is even more pronounced when it comes to
estimating the average velocity: 0.01~cm$/$s with the combined method in
comparison to 0.59~cm$/$s with local polynomials---next best of the
smoothing methods.

To generate simulated paths, the following steps are taken:
\begin{enumerate}[1.]
\item[1.] A pool of velocity profiles of different lengths and shapes is
generated.

\item[2.] At each step, a velocity profile is chosen at random from the
pool, and the length of the arrest following the progression is chosen
at random. The two are chained to the velocity profile. The total length
is larger than 30,000 records.

\item[3.] The true location is computed using the time series of
velocities (location at time 0 is at 0).

\item[4.] Independent $N(0,\sigma^{2})$ noise is added to the location
data.

\item[5.] 4\% of the nonarrests locations are chosen at random, and
their locations are shifted by 5, 10 or 15~cm (with equal probability)
to create outliers.

\item[6.] All locations are rounded to the nearest integer to reflect
the grid structure of the data.

\end{enumerate}

The above is repeated 50 times to generate replications of the paths for
each set of parameters. The following properties were computed for each
path:
\begin{enumerate}[1.]
\item[1.] The actual distance traveled using the sequence obtained at
stage 3 above. We denote the distance of the $i$th repetition
by $\theta_{i}$.

\item[2.] The estimated distance traveled using no smoothing.

\item[3.] The estimated distance traveled after smoothing using either
LOWESS, RRM or the combined method. The bandwidths used are the same as
used for real tracked data. We denote the estimated distances
by $\hat{\theta} _{i}$.

\item[4.] The true proportion of arrest time (0 velocity) is computed
from the velocity profile and denoted by $p_{i}$.

\item[5.] The estimated proportion of arrests is computed with no
smoothing and with each of the 3 smoothing methods to
obtain $\hat{p}_{i}$.

\end{enumerate}

We first simulated 100 paths of anesthetized mice, where the velocity
profile was a time sequence of 0, and no outliers were added (since
outliers occur mostly during progression). Table~\ref{table1} summarizes the
average and SD of distance traveled for these 100 simulated paths. The
averages are of the same order of magnitude as exhibited for the tracked
(real) anesthetized mouse.

\begin{table}
\caption{Average and SD of distance traveled over 100 simulated paths of an anesthetized mouse}\label{table1}
\begin{tabular*}{\textwidth}{@{\extracolsep{\fill}}ld{3.2}d{2.2}d{2.2}d{1.2}@{}}
\hline
 & \multicolumn{1}{c}{\textbf{Raw}} & \multicolumn{1}{c}{\textbf{LOWESS}} & \multicolumn{1}{c}{\textbf{RRM}} & \multicolumn{1}{c@{}}{\textbf{Combined}}\\
 \hline
Ave & 113.9 & 10.1 & 24.2 & 0.96\\
SD & 0.97 & 0.12 & 0.41 & 0.04\\
\hline
\end{tabular*}\vspace*{-5pt}
\end{table}

When choosing at random a velocity profile, the 50 repetitions have a
different underlying velocity profile and hence different distance
travel and proportion of arrest time. We define the following MSE as our
measure of performance:
\begin{eqnarray*}
\operatorname{MSE}(\theta) = \frac{\sum (\theta _{i} -\hat{\theta}_{i})^{2}}{50},\qquad
\operatorname{MSE}(p) = \frac{\sum (p_{i} - \hat{p}_{i})^{2}} {50} .
\end{eqnarray*}

Table~\ref{table2} summarizes the results of the true and estimated distanced
traveled as well as the MSE for the simulated paths with a velocity
profile that is not identically 0.

\begin{table}
\caption{True (simulated) distance traveled vs. estimated distance
traveled using raw  data,  LOWESS, RRM and the combined method}\label{table2}
\begin{tabular*}{\textwidth}{@{\extracolsep{4in minus 4in}}ll d{6.0}d{6.0}d{6.0}d{6.0}d{6.0}@{\hspace*{-3pt}}}
\hline
\multicolumn{2}{@{}l}{$\bolds{\sigma}$\textbf{:}} & \multicolumn{3}{c}{\textbf{0.6}} & \multicolumn{1}{c}{\textbf{1}} & \multicolumn{1}{c@{}}{\textbf{0.4}} \\[-6pt]
& & \multicolumn{3}{c}{\hrulefill}\\
\multicolumn{2}{@{}l}{$\bolds{\bar{p}}$\textbf{:}} & \multicolumn{1}{c}{\textbf{0.36}}
& \multicolumn{1}{c}{\textbf{0.74}} & \multicolumn{1}{c}{\textbf{0.64}} & \multicolumn{1}{c}{\textbf{0.36}} & \multicolumn{1}{c@{}}{\textbf{0.34}}\\
\hline
$\theta_{i}$ & Ave & 732 & 299 & 416 & 712 & 745\\
 & SD & 92 & 69 & 83 & 84 & 82\\[3pt]
Raw & $\bar{\hat{\theta}} _{i}$ & 967 & 609 & 704 & 1019 & 948\\
 & MSE & 55{,}487 & 95{,}924 & 83{,}566 & 94{,}511 & 41{,}548\\[3pt]
LOWESS & $\bar{\hat{\theta}} _{i}$ & 737 & 311 & 426 & 721 & 749\\
 & MSE & 31 & 139 & 100 & 77 & 15\\[3pt]
RRM & $\bar{\hat{\theta}} _{i}$ & 741 & 320 & 433 & 728 & 751\\
 & MSE & 89 & 419 & 294 & 263 & 34\\[3pt]
Combined & $\bar{\hat{\theta}} _{i}$ & 732 & 301 & 417 & 714 & 744\\
 & MSE & \multicolumn{1}{c}{0.07} & \multicolumn{1}{c}{3.1} & \multicolumn{1}{c}{1.6} & \multicolumn{1}{c}{5.5} & \multicolumn{1}{c@{}}{0.4}\\
\hline
\end{tabular*}\vspace*{-5pt}
\end{table}

Clearly, the combined method performs better than LOWESS and the RRM
separately. Although LOWESS is second to the combined method in
estimating the distance traveled, it fails to estimate the proportion of
arrest time, as is evident from Table~\ref{table3}, which shows the proportion of
arrest using each method and the corresponding MSE.

\begin{table}[b]
\caption{True (simulated) proportion of arrests vs. estimated proportion
using raw  data,  LOWESS, RRM and the combined method}\label{table3}
\begin{tabular*}{\textwidth}{@{\extracolsep{\fill}}lld{1.4}d{1.4}d{1.4}d{1.4}d{1.4}@{}}
\hline
\multicolumn{2}{@{}l}{$\bolds{\sigma}$\textbf{:}} & \multicolumn{3}{c}{\textbf{0.6}} & \multicolumn{1}{c}{\textbf{1}} & \multicolumn{1}{c@{}}{\textbf{0.4}} \\[-6pt]
& & \multicolumn{3}{c}{\hrulefill}\\
\multicolumn{2}{@{}l}{$\bolds{\bar{p}}$\textbf{:}} & \multicolumn{1}{c}{\textbf{0.36}}
& \multicolumn{1}{c}{\textbf{0.74}} & \multicolumn{1}{c}{\textbf{0.64}} & \multicolumn{1}{c}{\textbf{0.36}} & \multicolumn{1}{c@{}}{\textbf{0.34}}\\
\hline
Raw & $\bar{\hat{p}}_{i}$ & 0.20 & 0.33 & 0.29 & 0.16 & 0.23\\
 & MSE & 0.03 & 0.17 & 0.12 & 0.04 & 0.01\\[3pt]
LOWESS & $\bar{\hat{p}}_{i}$ & 0 & 0 & 0 & 0 & 0\\
 & MSE & 0.13 & 0.55 & 0.41 & 0.13 & 0.12\\[3pt]
RRM & $\bar{\hat{p}}_{i}$ & 0.41 & 0.72 & 0.64 & 0.41 & 0.39\\
 & MSE & 0.0026 & 0.0006 & 0.0001 & 0.0027 & 0.0022\\[3pt]
Combined & $\bar{\hat{p}}_{i}$ & 0.33 & 0.68 & 0.58 & 0.31 & 0.33\\
 & MSE & 0.0006 & 0.004 & 0.0027 & 0.0032 & 0.0001\\
\hline
\end{tabular*}
\end{table}

To summarize, the combined method is best in both aspects of estimating
distance traveled and proportion of arrest time. Using only LOWESS or
the repeated running median might be sufficient for one task but not for
both.

\section{Boundary and center estimation of an almost circular arena}\label{sec:3}

\begin{figure}

\includegraphics{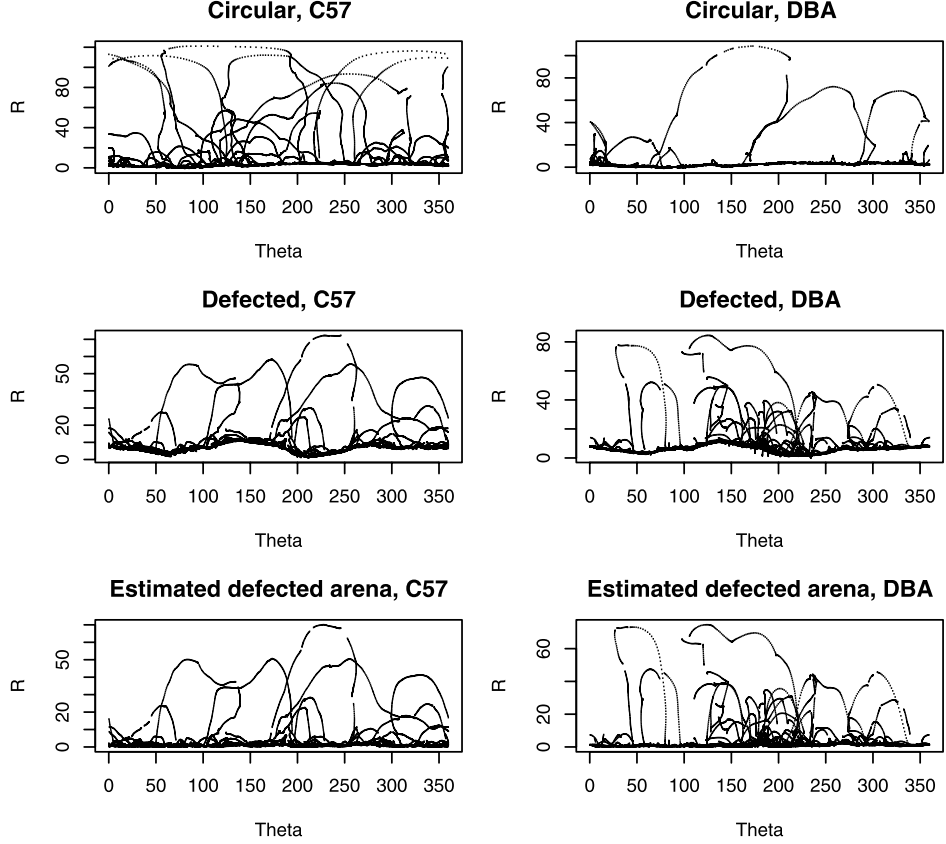}

\caption{Distance from a perfect circular wall versus angle
for two mice in the circular arena (top). The middle plots are the same
but for the distorted arena. The bottom plot show the distances from the
wall versus angle after boundary estimation.}\label{fig2}
\vspace*{-5pt}
\end{figure}

The wall of the arena is of major importance to the mouse, affecting its
behavior, in particular, the distance from the wall which is believed to
be related to anxiety [Hall (\citeyear{Hall1936}); Archer (\citeyear{Archer1973});
Walsh and Cummins (\citeyear{WalshCummins1976});
Finn, Rutledge-Gorman and Crabbe (\citeyear{FinnRutledgegormanCrabbe2003})]. In a perfect circular arena with known
radius and center, the distance is directly computed from the distance
to the center. In practice, an arena might have some deviations from a
perfect circle (sometimes hardly noticeable to the eye). The effect of
such deviations on distance from wall, if computed under the assumed
perfect circle, might be devastating. Figure~\ref{fig2} demonstrates the problem:
the distance from wall versus the angle throughout a session is plotted
for four mice: two from each strain (DBA and C57), two from each of two
laboratories. In the top plots the arena was indeed a circle, and the
large concentration of points near 0 was due to motion along the wall.
In the middle plots the arena was of a slightly distorted circular
shape. Watching the corresponding videotapes shows that these mice
tended to run along the wall in the same manner as the mice in the
circular arena. However, distance computations produced a wavy line
since distances were computed assuming a perfect circle. To correct
this, the distance between current location and actual boundary should
be measured. Unfortunately, such information is not available from the
tracking system.

One solution would be to rebuild the arena and rerun the experiment.
However, assuring perfect circularity is difficult and, furthermore, it
would not solve imperfect circularity in other laboratories running the
experiment. We were looking for a statistical solution that would
enhance the replicability of results across laboratories in future
studies.

Our solution is to estimate the actual boundary from the smoothed
location data of the mice. A key fact to the solution is that when a
mouse progresses along the wall it typically touches it. Hence, the
boundary can be inferred, indirectly, from the mouse's extreme
locations, as described below. The bottom two plots in Figure~\ref{fig2} show the
distance from the boundary in the distorted arena, after estimation.

In our situation, using a behavioral data form within the arena to
estimate its boundary is a necessity since data on actual boundary
locations is not available. However, even if measurements on the
boundary are available, obviously with noise, using the data from within
the arena might have statistical advantages when the latter is larger in
sample size (see Section~\ref{sec:3.5}).

\subsection{Estimation of the boundary of the arena}\label{sec:3.1}

Let us first assume that the location of the center of the arena is
known, so let it be at the origin. Let $(\tilde{x}_{i},\tilde{y}_{i})$
be the smoothed location at time $i$, and let $R_{i} =
\sqrt{\tilde{x}_{i}^{2} + \tilde{y}_{i}^{2}}$ and $\theta_{i}$ be its
polar representation. In the case of a perfect circular arena with
unknown radius, a natural estimate of the radius would be the maximum
observed distance.

When the circle is not perfect the distance between the wall and the
center is not constant, but may be assumed to change smoothly with the
angle. This motivates estimating the boundary using regression of
maximum distance on the angle. Some strains of mice tend to jump on the
wall (in particular, during lingering, but not only), and the location
of a jump is translated into locations outside the arena, hence, it is
better to use some high quantile of distance, rather than the maximum.
Thus, our problem can be phrased as that of a regression quantile of
$R_{i}$ on $\theta_{i}$, for a high quantile, and, in particular, its
nonparametric version to allow for local changes in the shape. The
resultant $\hat{R}_{p}(\theta)$ for $0 \le\theta\le 2\pi$ is the
estimated boundary. The regression quantile was first introduced in
Koenker and Bassett (\citeyear{KoenkerBassett1978}), and later extended to allow for a
nonparametric regression quantile [e.g., Koenker (\citeyear{Koenker2005})].

In principle, the quantile, $p$, might be different for different
strains. Currently, the maximum or high quantile can be used and the
results are almost identical. We used the 95th quantile for the
results presented here, but no visual differences were noticeable when
using the 99th quantile or even a higher one. An algorithm to
choose the quantile can be added for a fully automated procedure, but
until now there was no need for it.

The algorithm is limited to estimating only portions of the boundary
where behavioral data exist. In our case, this was not a problem.
However, two possible solutions are interpolation (since the boundary is
almost a circle) or using the boundary estimated from another mouse that
was recorded in the same arena.

\subsection{Quick and easy nonparametric regression quantile}\label{sec:3.2}

Implementation of a nonparametric regression quantile is involved and
requires sophisticated algorithms [Koenker (\citeyear{Koenker2005})]. Two different
approaches were taken by Koenker (and implemented in R in the package
``quantreg'') and by Yu and Jones (\citeyear{YuJones1998}). Using ``quantreg,'' we faced
several difficulties: convergence problems (not solved by
perturbations), slow execution time
and problem with large data sets
(with 30,000 locations the function did not run at all). We believe that
nonstatisticians, who are the target users of the proposed approach,
would be intimidated by such difficulties. We have therefore developed
an alternative, fast algorithm that uses the existing LOWESS algorithm.

The input is in polar representation of all smoothed locations during
progressions:
\begin{enumerate}
\item Divide the circle into $S$ sectors of angle $\Delta$, and let
$\alpha_{s}$ be the mid-angle of each sector for $s = 1,2,\ldots,S$.

\item Let $S_{s} = \{ (R_{k},\theta_{k})|\alpha_{s} - \Delta /2
\le\theta_{k} \le\alpha_{s} + \Delta /2\}$ be the collection of all
polar representation of locations within a sector.

\item For $s = 1,2,\ldots,S$, let $R_{s,p}$ be the $p$th quantile of
$\{R_{k}|(R_{k},\theta_{k}) \in S_{s}\}$.

\item Expand the data to produce an overlap at angles 0 and $2\pi$. This is
done by duplicating the beginning of the series at the end, and its end
at the beginning.

\item Regress $R_{s,p}$ on $\alpha_{s}$ on the expanded data, using LOWESS.
Use the estimated curve between 0 and $2\pi$ as the boundary estimate.
\end{enumerate}

The algorithm has two smoothing parameters. The size of a sector
$\Delta$ is the first one. We used $S = 720$ overlapping sectors
with $\Delta = 2\pi /360$. Experimentation with other numbers of sectors
did not reveal significant changes. Note that the bandwidth cannot be
too small (since it reflects a real boundary) or too large (since the
dents would have been noticeable to the eye). We used linear LOWESS
since in a small sector (1 degree amount to about 2~cm) the changes
cannot be too rough. The 2nd choice of a smoothing parameter is
the bandwidth of 0.15 for LOWESS. This was found in an iterative manner
while checking that the resultant curve was not too rough.

The biological implications of using the algorithm may be found in
Lipkind et al. (\citeyear{LipkindEtAl2004}).

\subsection{Estimations of the center of the arena}\label{sec:3.3}

\begin{figure}

\includegraphics{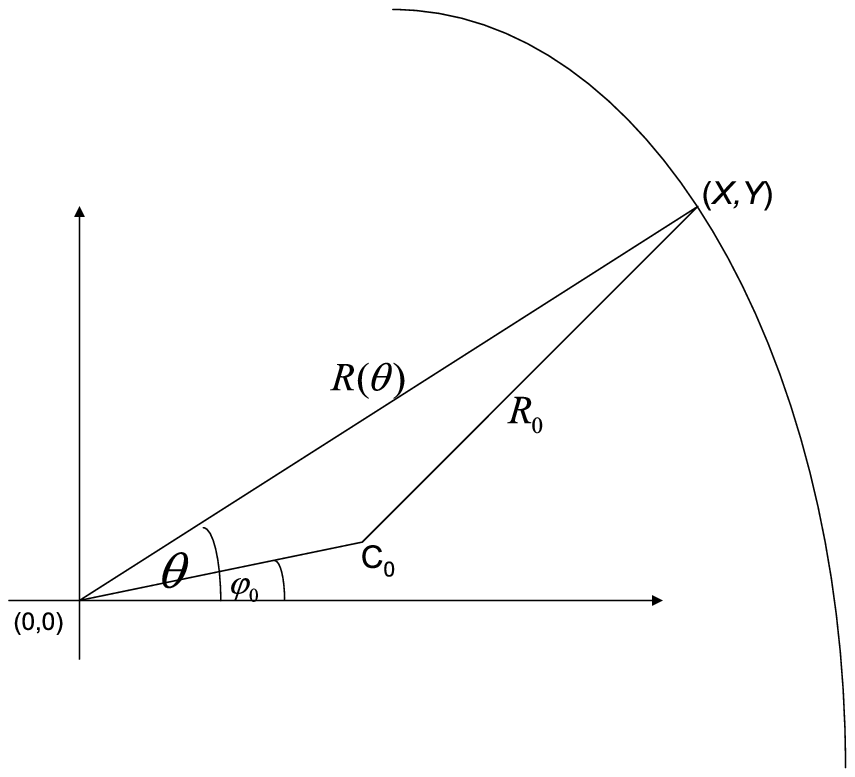}

\caption{The relations and angles between a point on the
boundary, the origin and the true center of the circle.}\label{fig3}
\end{figure}

So far, we have assumed the center of the arena is known. Now, assume
that the center $C_{0}$ is unknown, yet it is close to the origin. In
this case it can be estimated using the boundary.

See Figure~\ref{fig3} for clarification of notation. Let $(x,y)$ be a point on
the boundary, and denote its distance from $C_{0}$ by $R_{0}$. Let
$(R(\theta),\theta)$ be the polar representation of $(x,y)$ and
$(r_{0},\varphi_{0})$ be the polar representation of $C_{0}$. From the
cosine theorem it follows that
\[
R_{0}^{2} = R^{2}(\theta) + r_{0}^{2} - 2R(\theta)r_{0}\cos(\theta -
\varphi_{0}).
\]
Hence,
\[
R(\theta) = r_{0}\cos(\theta - \varphi_{0}) \pm R_{0}\sqrt{1 - ( r_{0}/ R_{0} )^{2}\sin ^{2}(\theta - \varphi _{0})}.
\]
By assumption, $r_{0}$ is small, so, using the Taylor approximation,
\begin{eqnarray*}
R(\theta)
&=& r_{0}\cos(\theta - \varphi_{0}) + R_{0} + o( r_{0}/R_{0} )
\\
&=& R_{0} + r_{0}\cos(\varphi_{0})\cos(\theta) +r_{0}\sin(\varphi_{0})\sin(\theta) + \varepsilon.
\end{eqnarray*}
In the last equation, $R_{0}$, $r_{0}$, $\varphi_{0}$ are unknown, while
$R(\theta)$, $\theta$ are known. There are many points along the boundary,
with polar representation: $R_{i}$, $\theta_{i}$. Using OLS, the parameters
$r_{0}$, $\varphi_{0}$ may be estimated from the boundary.

\begin{figure}

\includegraphics{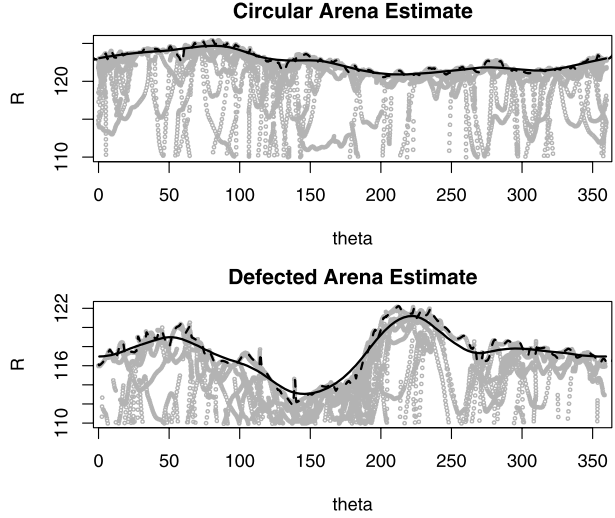}

\caption{Estimated arena wall versus angle using our algorithm
(solid) and ``quantreg'' (dashed). The grey points are the distances
from center for smoothed locations.}\label{fig4}
\end{figure}
%
\begin{figure}

\includegraphics{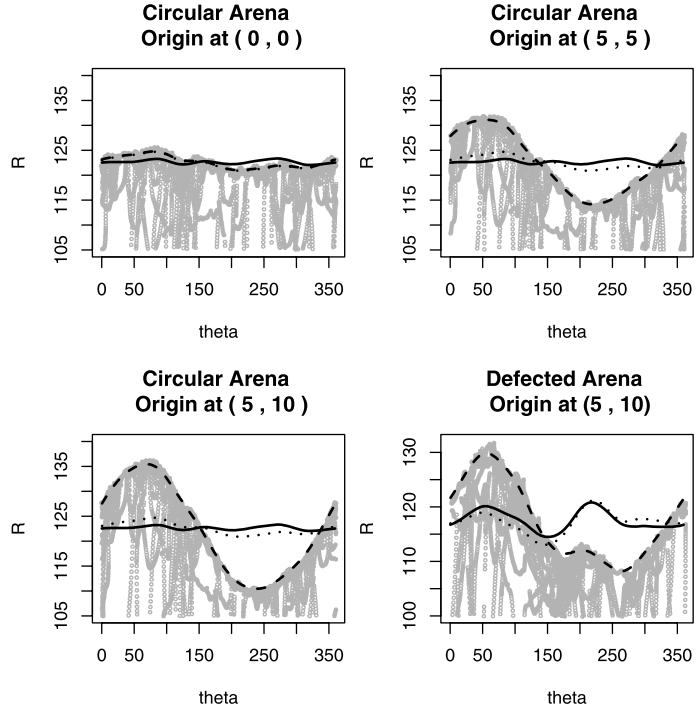}

\caption{Estimated arena wall versus angle when the center is
shifted. The grey points are as in Figure~\protect\ref{fig4}. The solid line is the
boundary estimate using our algorithm with center estimation. The dashed
line is our algorithm with no center estimation and the dotted line is
the our algorithm on the original data (i.e., center not shifted).}\label{fig5}
\end{figure}

In practice, the estimated boundary is used to estimate the center as
follows:
\begin{enumerate}[1.]
\item[1.] Use OLS to estimate $R_{0}$, $\beta_{1}$, $\beta_{2}$ in the
model $R_{i} = R_{0} + \beta_{1}\cos(\theta_{i}) + \beta_{2}\sin(\theta_{i}) + \varepsilon_{i}$.

\item[2.] Let $\hat{r}_{0} = \sqrt{\beta _{1}^{2} + \beta _{2}^{2}}$
and $\hat{\varphi} _{0} = \cos^{ - 1}(\beta_{1}/\hat{r}_{0})$.

\item[3.] Let, $\hat{x}_{0} = \hat{r}_{0}\cos(\hat{\varphi} _{0})$
and $\hat{y}_{0} = \hat{r}_{0}\sin(\hat{\varphi} _{0})$.

\end{enumerate}

\subsection{Advantages and evaluation of the algorithms to estimate
boundary and center}\label{sec:3.4}

The proposed algorithm to estimate the boundary is simple and,
consequently, it runs fast and converges well (which is especially
important as part of a high-throughput environment). Depending on the
size of the data, it runs 15--50~times faster than ``quantreg,'' and,
unlike ``quantreg,'' we have not encountered convergence problems. Its
disadvantage is the need to provide two bandwidths while ``quantreg''
requires only one. Cross-validation may be used to address this, but, in
practice, there was no need to do so.

Figure~\ref{fig4} compares the two methods. The grey points are the distances of
smoothed locations from $(0,0)$. The estimated boundary using our
algorithm is in the solid line and using ``quantreg'' is in the dashed
line. This was repeated for the circular and distorted arenas.
Qualitatively, the results are similar; however, ``quantreg'' seems too
rough for a physical boundary. When using other bandwidths ``quantreg''
did not converge.

\begin{figure}

\includegraphics{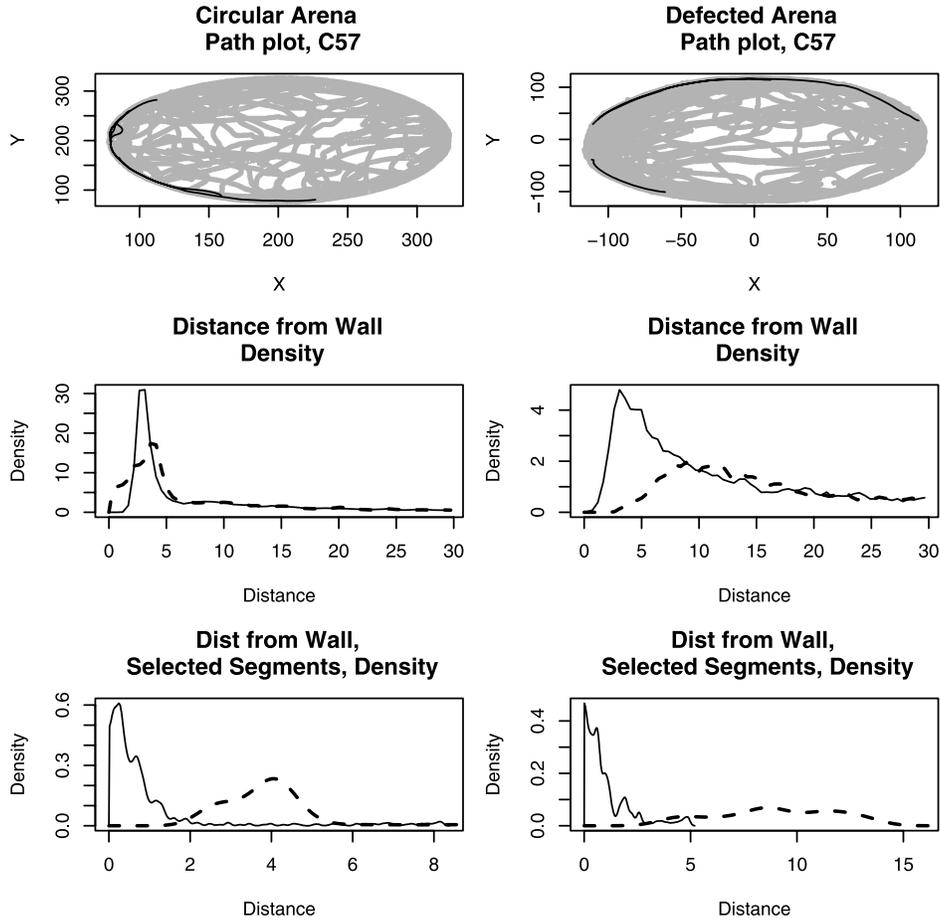}

\caption{Typical path plot of a C57 mouse in the circular and
distorted arenas. A few segments of movements along the wall were
selected and marked in black. The lower 4 plots show the densities of
distance from the wall for all points in the top path plot (middle) and
for the selected segments (bottom) in the two arenas. The solid line is
for the case where the distance is computed after boundary estimation,
while the dashed line is for the case where the distance is computed
assuming a perfect circle.}\label{fig6}
\end{figure}

Figure~\ref{fig5} demonstrates our algorithm with and without center estimation.
In the top left plot, data is from the circular arena and the presumed
center is $(0,0)$. The solid line is the estimated boundary when the
center was estimated as well. The dotted line is the estimated boundary
with no center estimation. The difference between the two is probably
because the center is not exactly at $(0,0)$. The top-right and
bottom-left plots are based on the same data, but the center is shifted
to the point mark at the title. The grey points are the distances from
$(0,0)$ and not from the true center. The solid and dashed lines are as
before. The bottom-right plot is the same but for the distorted arena.

Next we examine the density of distances from the boundary when the
algorithm to estimate the boundary is being used and when it is not. The
results are presented in Figure~\ref{fig6}. For both cases, all smoothed
locations within progression segments of a C57 mouse were taken. The
left plots correspond to the circular arena and the right plots to the
distorted one. The top plots show all the places in the arena that the
mouse visited at least once. The shape of the arena is not given, but it
can be deduced knowing that the mouse touches the boundary. For the
circular arena, five progression segments along the boundary were
chosen, and for the distorted arena three. These segments are marked in
black (with some overlap between them). The middle plots are the density
estimate of distances for all points belonging to progression segments.
The solid line is the distance from the estimated wall, using our
algorithm with center correction,
while the dashed line is the density
when the distances are taken from the assumed perfect circle. When the
arena is indeed a circle, the two are similar, but this is not the case
for the distorted arena. This effect is more dramatic when examining
only the selected segments that are performed close to the wall (bottom
plot). Here the effect of computing the distance from the wall, assuming
a perfectly circular wall, is evident.

\subsection{MSE comparison of boundary estimation}\label{sec:3.5}

The tracking system does not provide measurements of the boundary, so we
had to estimate it using behavioral data. Here, we demonstrate that even
if measurements along the boundary are available, obviously with noise,
using data within the arena might have statistical advantages in terms
of MSE due to the different sample sizes.

We assume the center is at the origin and the constant radius is
unknown, and compare estimation of the radius using boundary or
behavioral data. If the arena is not a perfect circle, a nonparametric
regression may be used to estimate the boundary (as described in Section~\ref{sec:3.1}).

Assume there are $n$ location measurements of the boundary and for each
the distance to the origin is computed so that $R_{i} = R + \varepsilon$,
where $\varepsilon$ have 0 mean and constant variance $\sigma^{2}$. Using
the mean to estimate $R$, the MSE is $\sigma^{2}/n$.

Alternatively, consider the location measurements during a session, and
assume that in each of the $n$ sectors there are $N$ measurements:
$Z_{ij}$ for $1 \le i \le n$ and $1 \le j \le N$, where $Z_{ij}$ are the
distances from the origin and have some distribution on a disk whose
center is at the origin and its radius is $R$. The MLE of
$R$ is $\max(Z_{ij})$.
\begin{lemma}
Assume $Z_{ij}$ are uniformly distributed on $[0,R]$ and let $\hat{R}_{1}
= \max(Z_{ij})$ and $\hat{R}_{2} = (nN + 1)\hat{R}_{1}/(nN)$. Then,
\begin{eqnarray*}
\operatorname{MSE} ( \hat{R}_{1}  ) &=& R^{2}\frac{2}{(nN +1)(nN + 2)}, \\
\operatorname{MSE} ( \hat{R}_{2}  ) &=& R^{2}\frac{1}{nN(nN +2)}.
\end{eqnarray*}
\end{lemma}

\begin{pf}
Calculating the pdf of $\hat{R}_{1}$ is straightforward and,
consequently,
\[
E( \hat{R}_{1} ) = \frac{nN}{nN + 1}R,\qquad
\operatorname{var}(\hat{R}_{1} ) = R^{2}\frac{nN}{(nN + 1)^{2}(nN + 2)} .
\]

The MSE of $\hat{R}_{1}$ and $\hat{R}_{2}$ follows easily.
\end{pf}

In our setup, the mice tend to stay near the boundary for a large
proportion of the time, hence, we consider a skewed distribution.
\begin{lemma}
Assume $Z_{ij} = RU_{ij}$ where $f(u) = (p + 1)u^{p}$ for $0 \le u \le
1$ and $p > 1$. Let $\tilde{R}_{1} = \max(Z_{ij})$ and $\tilde{R}_{2} =
\tilde{R}_{1}[nN(p + 1) + 1]/[nN(p + 1)]$. Then,
\begin{eqnarray*}
\operatorname{MSE} ( \tilde{R}_{1}  ) &=& R^{2}\frac{2}{ [nN(p + 1) + 1  ] [ nN(p + 1) + 2  ]}, \\
\operatorname{MSE} (\tilde{R}_{2}  ) &=& R^{2}\frac{1}{nN(p + 1) [ nN(p + 1) + 2 ]}.
\end{eqnarray*}
\end{lemma}

\begin{pf}
Calculating the pdf of $\tilde{R}_{1}$ is straightforward and,
consequently,
\begin{eqnarray*}
E( \tilde{R}_{1} ) &=& R\frac{nN(p + 1)}{nN(p + 1) + 1},
\\
\operatorname{var}( \tilde{R}_{1} ) &=& R^{2}\frac{nN(p + 1)}{[ nN(p + 1) + 1 ]^{2}[nN(p + 1) + 2 ]} .
\end{eqnarray*}
The MSE of $\tilde{R}_{1}$ and $\tilde{R}_{2}$ follows easily.
\end{pf}

The MSE based on the mean is $O(n^{ - 1})$, while the MSE based on the
MLE (or its unbiased version) of the behavioral data is $O(n^{ - 2}N^{ -
2})$. For the\vspace*{1.5pt} case of behavioral data, using $\hat{R}_{2}$, for example,
is advantageous over the boundary measurements if
\[
\frac{R^{2}}{\sigma ^{2}} < N(nN + 2).
\]
Similar comparisons are possible for the other estimators.

\section{Discussion}

For the dual purpose of smoothing locations and identifying arrests, we
combine LOWESS and the RRM. Using LOWESS echoes the approach of
Ramsay and Silverman (\citeyear{RamsaySilverman1997}), in which the path is viewed as a smooth location
function of time, and making use of its derivatives. Viewing the
resultant smoothed path as a very long paragraph with no punctuation,
the RRM adds the missing punctuation marks which, in turn, allows for
the analysis of each sentence. The idea of adding the punctuation marks
into the studied functional may be viewed as an extension of the
approach of Ramsay and Silverman.

Robustness in its traditional sense is an essential component in the
design of an automated high-throughput data analysis system, because it
automatically protects the analysis from sources of errors that could be
identified as gross errors once looked into by the human observer; alas,
this observer is missing from the initial stages of the high-throughput
process. A similar phenomenon happens in any data-mining operation, at
the stage of ware-housing the database, preparing it for further
analysis by sophisticated models and algorithms. The preparatory step is
always essential and automated, and the damage that can be done at this
stage is large. The use of classical robust procedures may need
adaptation, and the use of shortcuts to make the extra computational
effort feasible may be needed, as demonstrated by the examples given.
Such an emphasis on robustness when analyzing large data sets is not
usual, as robustness is associated with medium sized samples where the
gain in efficiency from using robust methods may be crucial.

Estimating the boundary provides protection from deviations from the
experimental design in our setup. Such deviations may happen, and when
the data are processed automatically, the methods used must be robust to
cope with them. This was achieved using a nonparametric regression
quantile. This approach extends the common practice in image processing
in which constant radius is assumed and estimated. Moreover, it turns
out that our solution is more flexible than we planned: initial
experimentation with the same algorithm to estimate the boundary of a
squared arena performed reasonably well, indicating that extending the
algorithm to take into account the possibility of corners at the
boundary will yield a good general solution.

Throughout the paper we have mentioned different choices made for
smoothing parameters. In all cases, the choices were made by an
iterative work of biologists and statisticians, and comparison of the
results to the video recordings themselves. The smoothing parameters are
potentially affected by arena size, animal size, recording rate and
height of ceiling, and their values should be fixated in the study
protocol. Alternatively, they can be estimated via some automatic method
such as cross validation based methods [e.g., Silverman (\citeyear{Silverman1986})], but the
algorithm should be fixated as well in the protocol and be identical for
all animals and groups involved. It is not clear to us whether the next
step should be a development of a more sophisticated method, driven by
data only, to choose the smoothing parameters or modeling the choice
made by an expert as a function of the parameters defined in the study
protocol (e.g., arena size, etc.). See, for example, the experience of
Likhvar and Honda (\citeyear{LikhvarHonda2008}), who demonstrated the limitations of
generalized cross validation in analyzing multiple time series, where
the chosen smoothing parameters occasionally missed the known curve
shape.

Computation of summaries (e.g., total distance, average speed during
progression) for each mouse is performed on the smoothed data, and is
followed by the assessment of differences between (inbred) strains. In a
single laboratory study this is done using the one-way ANOVA. Crabbe et
al. executed their study in 3 laboratories, and analyzed the data using
the two-way fixed ANOVA model, with strain and laboratory being the two
factors. They found the interaction to be significant and their
conclusion was the inability to declare replicability. In our view, the
mixed model, with laboratory and interaction being random, is more
appropriate [Kafkafi et al. (\citeyear{KafkafiEtAl2005})]. The mixed model is more
conservative than the fixed model, nevertheless, all 17 measures used in
Kafkafi et al. showed significant differences between strains (after
adjusting for multiplicity). We believe that smoothing and the ability
to create homogenous classes of behavior are crucial in achieving this,
and for that purpose the combination of LOWESS and the RRM plays a
central rule.

We hope that publishing this paper in a statistical journal will expose
more statisticians to the challenges in the field. There are open
problems both in connection with the current work and more generally in
the field of behavior genetics. Tracking and boundary estimation
discussed here are only two of them, and are encountered as statistical
problems in other fields as well.

\section*{Acknowledgment}

The statistical solutions discussed in the paper are implemented within
a free high-throughput software tool called SEE\break
[\href{http://www.tau.ac.il/\textasciitilde ilan99}{www.tau.ac.il/\textasciitilde ilan99}, Drai
and Golani (\citeyear{DraiGolani2001})].
The authors would like to thank Roger Koenker for the help provided in
the implementation of ``quantreg.''

\printaddresses

\end{document}